# FROM ASSESSMENT TO PRACTICE: IMPLEMENTING THE AIAS FRAMEWORK IN EFL TEACHING AND LEARNING

Jasper Roe [1], Mike Perkins [2] [*], Leon Furze

[1]Durham University, United Kingdom.

[2] British University Vietnam, Vietnam.

[3] Deakin University, Australia

[*] Corresponding Author: mike.p@buv.edu.vn

January 2025

## Abstract

Recent advances in Generative AI (GenAI) are transforming multiple aspects of society, including education and foreign language learning. In the context of English as a Foreign Language (EFL), significant research has been conducted to investigate the applicability of GenAI as a learning aid and the potential negative impacts of new technologies. Critical questions remain about the future of AI, including whether improvements will continue at such a pace or stall and whether there is a true benefit to implementing GenAI in education, given the myriad costs and potential for negative impacts.

Apart from the ethical conundrums that GenAI presents in EFL education, there is growing consensus that learners and teachers must develop AI literacy skills to enable them to use and critically evaluate the purposes and outputs of these technologies. However, there are few formalised frameworks available to support the integration and development of AI literacy skills for EFL learners. In this article, we demonstrate how the use of a general, all-purposes framework (the AI Assessment Scale) can be tailored to the EFL writing and translation context, drawing on existing empirical research validating the scale and adaptations to other contexts, such as English for Academic Purposes. We begin by engaging with the literature regarding GenAI and EFL writing and translation, prior to explicating the use of three levels of the updated AIAS for structuring EFL writing instruction which promotes academic literacy and transparency and provides a clear framework for students and teachers.

## Introduction

English as a Foreign Language (EFL) instruction has historically demonstrated an aptitude for integrating new technologies into practice, and a willingness to experiment with techniques that may provide new ways to improve language acquisition. Prior to the advent and development of the current generation of advanced GenAI models, multiple technologies have already been used to benefit writing practices. These include AI-powered digital writing assistants such as Grammarly, which can help provide suggestions for textual development (Grammarly, 2022). Such tools have been shown to improve grammatical accuracy, confidence, and autonomy and enhance feedback quality (Barrot, 2020; ONeill & Russell, 2019; Thi & Nikolov, 2021). Other technologies include automated paraphrasing tools which can assist in rephrasing text (Roe & Perkins, 2022) as well as sophisticated machine translation applications. However, these technologies are being supplanted by single GenAI tools, such as ChatGPT, which can support language learners in a wide variety of EFL related tasks.

In the broader field of education, the use of GenAI has led to consternation and debate. Research has evolved on the specific nature of risks regarding GenAI and plagiarism, or academic integrity violations (Cotton et al., 2023; Dehouche, 2021; Eaton, 2023; Kumar & Mindzak, 2024; Perkins, 2023; Wilder et al., 2021). Institutions and policies are now showing a shift towards acceptance of the use of GenAI in specific uses for text construction (Perkins & Roe, 2023a, 2023b). The acceptance of certain ethical uses of AI to assist with the writing process is of value to EFL learners given that writing in a second language causes cognitive stress (Gayed et al., 2022). Despite the potential value of decreasing learner stress, significant existential risks have been attributed to GenAI. In addition to the culturally biased output produced by such models, which may affect the validity of scientific knowledge (Messeri & Crockett, 2024; Roe, 2024), there is concern that the significant demand and techno-optimism associated with GenAI, combined with limited oversight from regulatory bodies or governments, will lead to the rapid depletion of natural resources (Bashir et al., 2024).

While it is important to acknowledge these highly concerning impacts, educators must also be pragmatic and deal with the issues that are currently arising. That is, there is limited guidance on how to use AI in education; at the same time, many students and teachers are using freely available models without sufficient support and preparation. In the U.S., 20% of all students surveyed were using ChatGPT for schoolwork (Gottfried, 2023), and both students and educators have expressed concerns with a lack of understanding regarding the use of GenAI tools in a variety of educational practices (Roe, Perkins, & Ruelle, 2024). Consequently, there is an urgent need for clearer guidance and frameworks to help learners and teachers use these tools effectively and critically.

One of the frameworks introduced to support educators with this goal was the AI Assessment Scale. First introduced in 2023 (Perkins, Furze, et al., 2024), it has since been revised (Perkins, Roe, et al., 2024). We use this five-point scale in this study to demonstrate how the classroom use of GenAI tools can be ethically integrated in an English teaching context. To do this, we first examine the current literature on EFL instruction and GenAI tools to demonstrate the evolution of the perspectives of this technology. We then introduce the AIAS before presenting an analysis of how this tool can be used to facilitate the classroom use of GenAI in an English teaching context, with a specific focus on Levels 2, 3, and 4 as uniquely suitable for the practice of EFL learning and teaching.

## The Evolution and Integration of AI in EFL Education

As with other disciplines, AI has had a large impact on EFL teaching and learning (Alshumaimeri & Alshememry, 2024). As these technologies have developed, so too have stakeholders' assessments of what students need to function in an EFL context rife with digital tools. For this reason, there has been an increased focus among practitioners and researchers on promoting various forms of literacy, beginning with technological literacy and progressing through media, digital, and finally to AI literacy today (Fan & Zhang, 2024; Roe, Furze, & Perkins, 2024). When adapting the AIAS for EFL purposes and to support EFL instruction, it is important to first explore the evidence base for how, why, and when AI may support EFL learning and teaching.

There is a large and developing body of literature on AI and GenAI in EFL education. Systematic reviews have identified that most studies in this area focus on writing skills (Alshumaimeri & Alshememry, 2024), while comparatively less focus has been given to other macro-skills, such as reading, listening, and speaking (Lo et al., 2024). Additionally, most studies focus on one specific GenAI tool, ChatGPT, and on text rather than multimodal outputs, which is perhaps unsurprising given the first-mover advantage which ChatGPT holds. Despite this advantage, it is anticipated that future applications will explore other multimodal GenAI outputs and a range of other AI applications, such as NotebookLM, Claude, or Gemini. Furthermore, there is evidence that multiple AI tools can benefit student learning in an EFL context, including Paperpal, Quillbot, and WordTune (Marzuki et al., 2023). Furthermore, as synthetic media and deepfake applications advance in terms of ease of use and cost-effectiveness, more attention may be given to this area. Some have argued that the integration of AI into education is inevitable (Koraishi, 2023), although this is a hotly debated issue. However, there is widespread recognition that AI is not a wholly positive technology, leading to the common metaphor of AI as a 'double-edged sword' (Derakhshan & Ghiasvand, 2024).

Research has both postulated and validated several positive outcomes for learners regarding interventions related to GenAI tools in classroom settings. A meta-analysis of 64 studies demonstrated that ChatGPT interventions significantly enhanced academic performance, affective-motivational states, and higher-order thinking propensities while reducing mental effort, although it had no significant impact on self-efficacy (Deng et al., 2024). A review of 21 experimental and quasi-experimental studies found that EFL instruction with the use of AI chatbots was more effective than that without (Wu & Li, 2024). Among the many tasks that ChatGPT can achieve in relation to language learning, Kohnke et al. (2023) identified correcting and explaining mistakes, creating multiple genres of text, developing quizzes, giving definitions, and several more, indicating a wide and broad range of tasks that GenAI can contribute to. Similarly, other authors highlight that AI tools may benefit vocabulary acquisition, pronunciation, and student engagement (Algraini, 2024; Mohamed, 2024). Use of GenAI tools like ChatGPT may also benefit motivation for EFL learners (Huang & Mizumoto, 2024; Yuan & Liu, 2025), improve willingness to engage in conversation (Yang et al., 2022), boost overall engagement (Wang & Xue, 2024), improve students' abilities to enter into a 'flow state' (Zhang et al., 2021) and enable EFL students to set higher goals (Guo & Li, 2024) and express their ideas in English (Zhao, 2022) and collaborate (Teng, 2024). GenAI technologies may assist with language learning beyond the classroom through Informal Digital Learning of English (IDLE) (Liu et al., 2024), and visually embodied GenAI chatbots may enhance the emotional aspects of language learning using human-like avatars (Wang et al., 2024).

As AI-powered language learning tools may be able to adapt to individual needs (Mohamed, 2024), personalisation of learning materials, for example text reading passages adapted to learner profiles, is possible, reducing both the burden on material design for the teacher (Koraishi, 2023) and providing increased speed and quality of learner feedback (Dai & Liu, 2024). Research has also explored how GenAI and ChatGPT are viewed by EFL teachers, with Slamet (2024) finding that teachers express a positive perception of ChatGPT to benefit student access to linguistic resources, and Ulla et al. (2023) finding similar positive intentions to use ChatGPT for lesson planning and resource creation. In writing and feedback, the ChatGPT has been shown to have higher reliability coefficients for marking EFL work than human raters (Li et al., 2024).

**Understanding AI's undesirable implications for EFL learning and teaching**

Despite findings in EFL research demonstrating a multitude of positive benefits for learners across multiple skills and areas of language acquisition, there is a consensus in the literature that AI should not be implemented in an unstructured way, and tools such as chatbots can, in some cases, constrain rather than improve language learning (Jeon, 2022). Furthermore, the literature suggests that AI use requires teachers to engage in professional training to be prepared to embed AI into EFL where appropriate (Jiang, 2022), which must be done with the awareness that such technologies can provoke strong emotional reactions from teachers and learners (Shen & Guo, 2024; L. Yang & Zhao, 2024; Yin et al., 2024). Consequently, a critical approach to implementing GenAI in EFL practice is essential.

GenAI models, such as ChatGPT, may also lack cultural sensitivity (Werdiningsih et al., 2024) and present a worldview aligned with those in the training data, which by its nature focuses on Western cultural norms. The novel nature of AI chatbots may generate fear and anxiety among EFL learners (Yang & Zhao, 2024). Therefore, teachers must play a pivotal role in guiding technology in the classroom (Mohamed, 2024), while aiming to develop learners' critical literacy (Darwin et al., 2024). Additional concerns to be aware of when introducing AI to the EFL classroom include the possibility of impacting creativity and spreading false information (Derakhshan & Ghiasvand, 2024) as well as overreliance on the tools. (Darvishi et al., 2024; Gao et al., 2024; Yuan et al., 2024) or the dishonest usage of AI-powered writing tools (Roe, Renandya & Jacobs, 2023). Accessibility and inclusion are concerns, as free versions of GenAI tools have user limits when compared to the paid version of the same tools (Nizzolino, 2024); therefore, instructors must focus on providing equal learning opportunities (Kohnke, 2023) and mitigating the pressure for learners to subscribe to premium models (Yuan et al., 2024). These myriad challenges and risks highlight the importance of a structured, cautious approach for integrating AI into EFL.

**Introducing the AIAS**

The AI Assessment Scale (AIAS) is a flexible framework originally conceived in 2023 (Perkins, Furze, et al., 2024) and updated in 2024 (Furze, 2024a; Perkins, Roe, et al., 2024). The AIAS was designed as a simple and practical tool to help educators and students deal with the sudden emergence of GenAI tools and their astonishing capabilities in accomplishing a variety of assessment tasks. This was a result of the fact that a small proportion of students could use GenAI to subvert the validity of assessment and achieve a passing level without putting in the required effort and time themselves (Thompson et al., 2023), and a broader view that as disruptive technologies develop, assessment would have to fundamentally change. The view that assessment in higher education must be re-evaluated has now received widespread attention in the literature (Bearman et al., 2024; Mao et al., 2024; Rasul et al., 2024; Thanh et al., 2023; Thompson et al., 2023; Xia et al., 2024). This has become even more critical as the abilities of GenAI tools to tackle assessment have grown even further since their first release, including demonstrating the ability to pass high-stakes medical and legal examinations (Head & Willis, 2024; Newton et al., 2024; Newton & Xiromeriti, 2024).

The AIAS, in its original incarnation, offered a five-point scale that was intended to guide educators and students on how to use AI tools within assessments, while maintaining transparency and academic integrity, based on the principles of clear, simple, and two-way communication. This is important given the gap between learners and teachers' opinions on how to use AI (Smolansky et al., 2023). The lowest level of the AIAS was intended to cover tasks where security was vital and no external tools were allowed (No AI), while the fifth level (Full AI) was intended to allow learners to make full use of the affordances of the new technology to meet specific assessment goals. The AIAS has been adopted by universities and K-12 institutions, and promoted by governmental agencies as a potential tool to assist educators in dealing with the implications of GenAI (Lodge, 2024). It has been translated into multiple languages (Furze, 2024b) and modified significantly for the K-12 and higher educational contexts. Furthermore, empirical research has demonstrated that implementing the AIAS in a tertiary context can stimulate a shift in pedagogical practices on behalf of assessors and promote academic integrity (Furze et al., 2024a).

At the time of writing, the AIAS has been updated to replace the fourth level (AI and Human Evaluation) with the previous fifth level (Full AI) and the addition of a new level (AI Exploration). The rationale behind this change is that since the original framework was released, GenAI has become so commonplace and so advanced that there may now be a need to actively promote and even expect the use of such tools in assessment in an exploratory manner. To this end, Full AI (in which all use of AI tools is permitted) is now replaced with 'AI exploration' which may involve assessment design that necessitates engagement with GenAI to meet learning outcomes. Stylistically, changes were also made to the colour scheme to distinguish it from a 'traffic-light system' which may implicitly have assumptions that lower levels of the scale, previously coded red, suggested a negative connotation, and that higher levels of the scale had more positive connotations. The current version of the AIAS is shown in Figure 1.

*Figure 1. The AIAS*

## The AI Assessment Scale

| 1 | NO AI | The assessment is completed entirely without AI assistance in a controlled environment, ensuring that students rely solely on their existing knowledge, understanding, and skills<br>**You must not use AI at any point during the assessment. You must demonstrate your core skills and knowledge.** |
|---|---|---|
| 2 | AI PLANNING | AI may be used for pre-task activities such as brainstorming, outlining and initial research. This level focuses on the effective use of AI for planning, synthesis, and ideation, but assessments should emphasise the ability to refine and refine these ideas independently.<br>**You may use AI for planning, idea development, and research. Your final submission should show how you have developed and refined these ideas.** |
| 3 | AI COLLABORATION | AI may be used to help complete the task, including idea generation, drafting, feedback, and refinement. Students should critically evaluate and modify the AI suggested outputs, demonstrating their understanding.<br>**You may use AI to assist with specific tasks such as drafting text, refining and evaluating your work. You must critically evaluate and modify any AI-generated content you use.** |
| 4 | FULL AI | AI may be used to complete any elements of the task, with students directing AI to achieve the assessment goals. Assessments at this level may also require engagement with AI to achieve goals and solve problems.<br>**You may use AI extensively throughout your work either as you wish, or as specifically directed in your assessment. Focus on directing AI to achieve your goals while demonstrating your critical thinking.** |
| 5 | AI EXPLORATION | AI is used creatively to enhance problem-solving, generate novel insights, or develop innovative solutions to solve problems. Students and educators co-design assessments to explore creative AI applications within the field of study.<br>**You should use AI creatively to solve the task, potentially co-designing new approaches with your instructor.** |

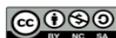 Perkins, Furze, Roe & MacVaugh (2024). The AI Assessment Scale

Despite being a tool primarily geared towards assisting with the redesign and redevelopment of assessment in the age of GenAI, the AIAS can equally be used as a framework for structuring course outcomes, learning activities, or pedagogical tasks. For example, the original AIAS has been adapted to provide guidance on how to develop learners English for Academic Purposes (EAP) skills by redefining the levels to align with common tasks found in an EAP context (Roe, Perkins, & Tregubova, 2024). Again, this approach demonstrates the flexibility of the AIAS as a tool which can be adapted to different contexts, yet supports a clear set of principles underpinning the integration of GenAI into educational assessment. These are that it must be built on a clear set of transparent expectations and rules, that it must be adaptable as technology changes, and that it must be critically and carefully implemented.

## The AIAS for EFL: Why and How?

Given the above, there are potential benefits to GenAI which can be used in EFL, not just for writing, but in contributing to a number of different skills and supporting many different tasks. To achieve this, it is important to cultivate AI literacy and facilitate informed engagement (Yan et al., 2024) through correctly scaffolded instructions. This relates to broader calls in EFL for teachers and students to be able to recognise the limitations of GenAI and use it safely and with integrity (Kohnke et al., 2023). As a result, we suggest that the AIAS can be adapted for in-class EFL use, not just for structuring assessments for and of learning, but also for broader task-based approaches. In particular, Levels 2, 3, and 4 have special applicability to common tasks in an EFL classroom context.

### AIAS Level 2 for EFL: AI Planning

Level 2 of the AIAS represents 'AI planning'. This level is particularly suitable for guiding common EFL tasks, including pre-task activities. For example, if planning a discussion task based on reading for a group of upper-intermediate learners, then Level 2 may be assigned. In this context, learners would not be able to use AI tools 'on-the-fly' during the group discussion but may be encouraged to use tools such as ChatGPT, Claude, or NotebookLM to help them understand the reading through clarification of content, support with explaining new vocabulary, or summarising key ideas. AI tools can be particularly beneficial and scaffold language learning in these tasks. At the same time, learners will need to be given adequate instruction on the limitations and possibilities of tools that generate inaccurate information. Learners will need to pay close attention to assessing the quality of the output, thus developing evaluative judgement, a key skill for the age of AI (Bearman et al., 2024).

A second example in which Level 2 of the AIAS may be beneficial for EFL instruction is written communication. In the AIAS adaptation for EAP, we argued that GenAI tools may help learners to become familiar with the drafting process and generating their own feedback on text that they have produced (Roe, Perkins, & Tregubova, 2024). In relation to EFL more generally, GenAI may be used in a Level 2 series of activities to provide feedback on written text production. This is an area where GenAI tools excel-providing quick, user-friendly feedback on how to improve the grammatical accuracy, style, flow, and coherence of written text, and explain the reasons behind these. Educators may also encourage the use of prompting techniques to encourage personalisation of the output – for example, asking the GenAI tool to make text easier to understand for an EFL learner. Depending on the first language of the learner, GenAI tools may even be able to have a dialogue in both languages, thus supporting translanguaging skills.

## AIAS Level 3 for EFL: AI Collaboration

AIAS Level 3 refers to 'AI Collaboration'. There is a strong alignment between EFL and the AIAS in this context. Although this collaboration between human and GenAI tool is targeted toward the completion of assessment tasks, it can also be extended and adapted to learning tasks in the EFL classroom. Indeed, a strong research base suggests that collaborative approaches are effective for implementing AI within an EFL context. For example, Lee (2024) highlights that using ChatGPT as a writing assistant collaboratively in EFL can help students feel reassured due to the ongoing, instantaneous feedback it can provide, while simultaneously helping to foster patience and involvement with the learning task. Furthermore, Guo and Wang (2024) asserted that a collaborative approach for teachers with ChatGPT can offer effective feedback for student writing. This approach is aligned with empirical studies that have shown that students are more open to AI-supported feedback from teachers than AI feedback alone (Roe, Perkins, & Ruelle, 2024).

For Level 3 focused activities, learners can be encouraged to co-create texts or even speech output (for multimodal GenAI tools) using the target language, phrases, or structures. For higher level learners, attention can be drawn to challenges sometimes found in GenAI output, such as 'wordy' or overly verbose responses, as well as hallucinated or fabricated content. (Kohnke et al., 2023). Such collaborative approaches have already been reported in literature, showing that such techniques can improve L2 writing proficiency (Wiboolyasarin et al., 2024) and academic writing proficiency (Nguyen et al., 2023). Collaborative tasks may also include group work or language games; for example, prompting an AI tool to generate quiz questions or resources for other students. Another possible application for the formative assessment of learning may be to ask learners to submit an unedited piece of writing for peer evaluation and then another subsequently edited with GenAI.

## AIAS Level 4 for EFL: Full AI

Level 4 of the AIAS (Full AI) encourages learners to make use of any available tools that they see fit when tackling an assessment task. Again, this experimental and all-inclusive approach may be beneficial at times in language classrooms, although it requires careful thought and planning to mitigate the known risks of AI technologies. These technologies can be used to support EFL learners in developing confident voices and styles in a new language. Level 4 activities could encourage users to see how effective an application like ChatGPT is for tackling a complex writing or reading test (encouraging the learners to 'test' the tool), or could involve asking learners to generate their own self-study resources using GenAI applications. Full AI tasks are useful for encouraging an experimental and creative approach to not only assessment, but also language learning in general. On the other hand, this should be initiated in a way that encourages students to be judicious and cautious in utilising new technologies and guiding them toward trusted applications. An example of a Full AI task may be for a group of learners to produce a poster, presentation slides, or other resources using whichever AI tools they wish to use (e.g. image, text, or entire presentation slide set generation). Assessments here focus on both linguistic elements of the output as well as the demonstration of more technical skills; however, this approach assumes the requisite level of AI literacy and familiarity with use. A Full AI task should only be explored after students have demonstrated that they have a clear understanding of the limitations, ethical considerations, and practical techniques for using GenAI in this setting.

EFL learners often spend more time on low-level writing tasks such as word choice and translation, thus giving less time to spend on higher-level structural tasks such as coherent organisation (Gayed et al., 2022); thus, assigning a Full AI task may encourage learners to cognitively offload such lower-order tasks and spend more time engaging with higher-order concerns. Simultaneously, for more proficient EFL learners, the critical evaluation of GenAI outputs can form part of a Full AI learning task. New 'speech-to-speech' models such as OpenAI's Advanced Voice Mode and Google's Gemini 2.0 Advanced Voice Mode have the capability to not only speak in multiple languages, but to respond idiomatically and with varied dialects. These technologies can be used to support EFL learners in developing confident voices and styles in a new language.

## Summary

All levels of the AIAS can be adapted to the assessment of learning in the GenAI era, and it is essential to have a balance of tasks – including the use of Level 1 'No AI' assessments where assurance of learning, and task validity are paramount. Simultaneously, using a structured approach to collaborate with students, foster AI literacy, and encourage open dialogue on the place that AI tools have in EFL education. From this perspective, we argue that Levels 2, 3, and 4 of the AIAS are the most suitable for such activities.

A summary of the different suggestions for implementation in the EFL context is presented in Table 1.

*Table 1 Summary of the Application of AISA Levels in EFL*

| AIAS Level | Adaptation to EFL Activities |
|---|---|
| Level 2: AI Planning | AI planning can be applied to a variety of EFL activities, not limited to assessments, including preparation for class discussions, quizzes, or knowledge acquisition on a topic in content and language integrated learning (CLIL). |
| Level 3: AI Collaboration | AI collaboration activities can include co-creation of texts, learning resources, and using GenAI tools as a critical friend or teacher. |
| Level 4: Full AI | AI tools can be used critically to support the learning activity, as a form of experimentation or a way of fostering AI literacy. Examples may include asking learners to develop their writing for different genres by asking an AI chatbot for help and advice, or finding answers to language-related questions. A Full AI classroom experience can help learners to understand the benefits and limitations of using new technology to support language learning. |

## Discussion

Adapting the AIAS to EFL contexts represents an opportunity to enhance the structure and guidance for both language teachers and students. The AIAS has been implemented successfully in higher education at both the institutional and individual levels (Furze et al., 2024b) and has shown promise in guiding the use of AI technology in assessment and learning. While all levels of the AIAS may be used for assessment throughout a course, learning activities that support language development may also apply the AIAS, providing a clear approach and scaffolding for learners' knowledge and experience with AI while developing critical digital literacy. Within the AIAS levels, we contend that Levels 2, 3, and 4 show particular promise for this purpose. The AIAS framework's graduated approach to AI integration may be particularly suited to younger generations, given their great familiarity with technology in studies and in their lives outside of formal education (Polakova & Ivenz, 2024).

However, there are concerns regarding unequal resource access, which could contribute to a digital divide. In deciding whether to implement an AIAS-type set of activities in the EFL classroom, the ultimate decision must be made contextually and with due consideration of equity, learner needs, and the overall approach to teaching; technology should be implemented judiciously and only when it is beneficial to overall learning. In our review of the GenAI and EFL literature, several authors reported the need for a collaborative approach with AI tools; for example, Kartal (2024) and Lee (2024) highlighted the value and need for collaboration, while Roe et al. (2024) noted that GenAI tools do not have to replace or detract from, but rather complement, learners' skills in areas such as paraphrase. Consequently, Level 3 of the AIAS may be one of the highest-potential forms of AI implementation in an EFL context to scaffold idea generation and language production.

The fact that the framework is flexible, customisable, and can be adapted to different tasks suggests that it may be able to effectively meet the needs of EFL learners. Rezai et al. (2024) asserted that AI's capacity to improve vocabulary, grammatical accuracy, and pronunciation while providing immediate feedback is relevant. These features are especially relevant to tasks introduced under Level 2 (AI Planning) and Level 3 (AI Collaboration) of the AIAS framework.

However, several important caveats should inform the implementation of the AIAS framework in EFL contexts. Mohamed's (2024) observation that technology cannot replace human interaction serves as a reminder that the framework should not be seen as a replacement for traditional language teaching and learning approaches, nor for pedagogies that have stood the test of time. Ultimately, we must take a careful, cautious, and critical approach to experimenting with new technologies in an educational context, and issues of equity, inclusion, and data privacy must be considered.

## Conclusion

Adapting the AIAS framework to EFL teaching and learning contexts is complex. We argue that a structured, scalar approach can benefit learners and educators by providing a clear and transparent method to ensure two-way communication on how to use new technologies for specific tasks (Furze et al. 2024b). Our analysis of the EFL literature has demonstrated key areas of alignment with the affordances of AI technology and EFL learning and the structure provided by Levels 2, 3, and 4 of the AIAS. This framework can help develop approaches to include AI in the classroom, without surrendering its use in every situation, and without completely banning such tools given their multiple affordances for learning. Simultaneously, learners may be able to develop AI literacy skills while benefiting from improved language acquisition in certain AIAS-informed tasks. The framework's graduated approach to AI integration allows for scaffolding of language skills and digital literacy.

When making the decision as to whether GenAI tools might be integrated into the EFL classroom, educators and institutions both have responsibilities for their learners. Individual educators have a responsibility in considering the equity and accessibility concerns of the class, the AI technologies to be used and their benefit, and the overarching concerns regarding such technologies (for example, their energy use, cultural biases, and potential inaccuracies, to name a few). Institutions bear the responsibility of ensuring that if AI is to be included in EFL education, educators will be provided with professional development to help develop their understanding and use of new and emerging technologies.

In the near future, we hope to see empirical research which explores the impacts of AIAS-structured activities on language acquisition, especially in areas such as writing development, vocabulary acquisition, and overall communicative competence. Research exploring such applications in a variety of cultural and institutional contexts could help uncover insights that would provide directions for future adaptations of the framework.

As AI technologies continue to expand and become embedded in daily life and educational contexts, simple and effective tools and frameworks that may help guide pedagogical practices will become increasingly important. While the AIAS, as a fundamentally assessment-oriented framework, was originally designed for higher education and K-12 contexts at large, there are clear areas of alignment with non-assessment-focused learning activities in the EFL context. Therefore, the AIAS may be worthy of consideration by educators and institutions as one piece of the puzzle in addressing how to use AI in EFL education.